\documentclass[aip,apl,reprint,superscriptaddress,longbibliography,floatfix]{revtex4-1}
\usepackage{graphicx}
\usepackage{siunitx}
\usepackage[utf8]{inputenc}
\usepackage{newunicodechar}
\usepackage{float}
\usepackage{amsmath}
\usepackage{xcolor}

\begin{document}

\title{Self-organized nonlinear gratings for ultrafast nanophotonics}

\newcommand{\NISTTF}{Time and Frequency Division, National Institute of Standards and Technology, Boulder, CO 80305, U.S.A.}
\newcommand{\NISTG} {Center for Nanoscale Science and Technology, NIST, Gaithersburg, MD 20899, U.S.A.}
\newcommand{\CU}    {Department of Physics, University of Colorado, Boulder, CO, 80309, U.S.A.}
\newcommand{\soft}{Department of Physics and Soft Materials Research Center, University of Colorado, Boulder, CO 80309, U.S.A.}
\newcommand{\ECEE}{Department of Electrical, Computer, and Energy Engineering, University of Colorado, Boulder, CO 80309, U.S.A.}
\newcommand{\RASEI}{Renewable and Sustainable Energy Institute, NREL and University of Colorado, Boulder, CO 80309, U.S.A.}
\newcommand{\hopkins}{Department of Electrical and Computer Engineering, Johns Hopkins University, Baltimore, MD 21218, U.S.A.}

\author{Daniel~D.~Hickstein} \email{danhickstein@gmail.com}
 					         \affiliation{\NISTTF}
\author{David~R.~Carlson}    \affiliation{\NISTTF}
\author{Haridas~Mundoor}     \affiliation{\soft}
\author{Jacob~B.~Khurgin}    \affiliation{\hopkins}
\author{Kartik~Srinivasan}   \affiliation{\NISTG} 
\author{Daron~Westly}        \affiliation{\NISTG} 
\author{Abijith~Kowligy}     \affiliation{\NISTTF}
\author{Ivan~Smalyukh}       \affiliation{\soft}
                             \affiliation{\ECEE}
                             \affiliation{\RASEI}
\author{Scott~A.~Diddams}    \affiliation{\NISTTF}
                             \affiliation{\CU} 
\author{Scott~B.~Papp}       \affiliation{\NISTTF}
                             \affiliation{\CU}
                             
\date{\today}

\renewcommand{\topfraction}{0.9}	
\renewcommand{\bottomfraction}{0.8}	
\setcounter{topnumber}{2}
\setcounter{bottomnumber}{2}
\setcounter{totalnumber}{4}     
\setcounter{dbltopnumber}{2}    
\renewcommand{\dbltopfraction}{0.9}	
\renewcommand{\textfraction}{0.07}	
\renewcommand{\floatpagefraction}{0.8}	
\renewcommand{\dblfloatpagefraction}{0.7}	

\begin{abstract}
Modern nonlinear optical materials allow light of one wavelength be efficiently converted into light at another wavelength. However, designing nonlinear optical materials to operate with ultrashort pulses is difficult, because it is necessary to match both the phase velocities and group velocities of the light. Here we show that self-organized nonlinear gratings can be formed with femtosecond pulses propagating through nanophotonic waveguides, providing simultaneous group-velocity matching and quasi-phase-matching for second harmonic generation. We record the first direct microscopy images of photo-induced nonlinear gratings, and demonstrate how these waveguides enable simultaneous $\chi^{(2)}$ and $\chi^{(3)}$ nonlinear processes, which we utilize to stabilize a laser frequency comb. Finally, we derive the equations that govern self-organized grating formation for femtosecond pulses and explain the crucial role of group-velocity matching.  In the future, such nanophotonic waveguides could enable scalable, reconfigurable nonlinear optical systems.
\end{abstract}

\maketitle

\section{Introduction}
Nonlinear light--matter interactions can convert photons to new frequencies, and serve as fundamental tools for laser science \cite{new2011}, telecommunication \cite{schneider2004}, quantum computation \cite{weston2016}, and other disciplines across science and technology. As these applications become increasingly commonplace \cite{garmire2013}, there is a need for nonlinear optical platforms that are small, easy to fabricate, low cost, and efficient. Achieving high conversion efficiency requires matching the phase velocities of each wavelength of light used in the nonlinear interaction. In some cases, a fortuitous material birefringence can provide a mechanism for this phase-velocity matching, but in many cases, quasi-phase-matching (QPM) techniques must be used instead. QPM typically involves periodically switching the nonlinearity of the material on the micrometer scale, however, such small features are often difficult to fabricate, and are only possible in certain materials. When femtosecond pulses are used for nonlinear optics, achieving the best performance becomes doubly challenging because, in addition to phase-velocity matching, \textit{group-velocity} matching must also be maintained. Meeting these two conditions simultaneously is usually difficult and failing to do so results in limited conversion efficiency or temporal broadening of the ultrashort pulses.

\begin{figure}
	\includegraphics[width=\linewidth]{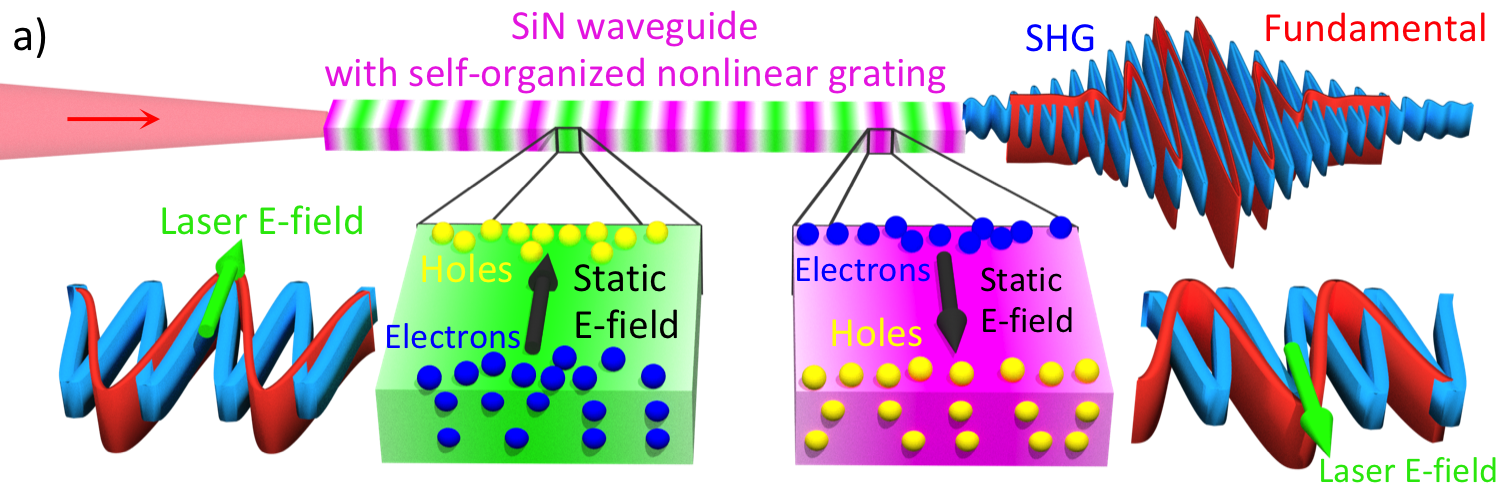}
  	\includegraphics[width=\linewidth]{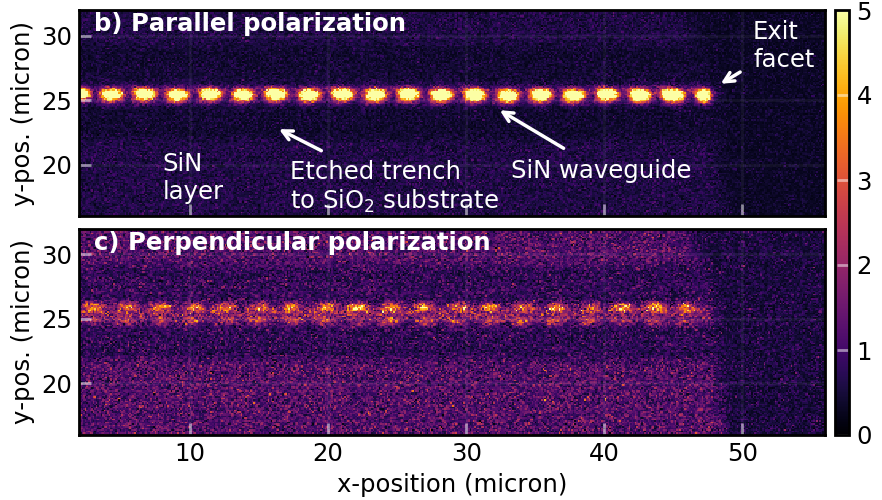}
	\caption{\label{fig:overview} \textbf{Self organized nonlinear grating (SONG) formation in a silicon nitride (SiN) nanophotonic waveguide.} a) The interference of fundamental and second-harmonic (SH) light forces positive charges to one side of the waveguide and negative charges to the other, forming a static electric field, enabling an effective $\chi^{(2)}$. The phase walk-off between the two fields switches the direction of the electric field, automatically generating a self-organized nonlinear grating with the correct periodicity to provide quasi-phase-matching for the SHG process. The geometry of the waveguide provides group-velocity matching, allowing for enhanced SH conversion efficiency. b) An second-harmonic-generation (SHG) microscopy image reveals the periodic modulation of the effective $\chi^{(2)}$. c) An SHG microscopy image of the same waveguide with perpendicular polarization reveals a double-lobed structure, confirming the charge localization.}
\end{figure}

To avoid the difficulties of directly fabricating QPM materials, it is possible the use laser light to generate ``self-organized nonlinear gratings'' (SONGs), which can provide QPM for nonlinear processes~\cite{dianov1994,balakirev1997,anderson1991,stolen1987}. For example, in materials such as $\mathrm{SiO_2}$ that do not normally exhibit a bulk quadratic nonlinearity ($\chi^{(2)}$), irradiation with a laser can form permanent electric fields in the material, which act on the material's cubic nonlinearity ($\chi^{(3)}$) to produce an effective quadratic nonlinearity $\chi^{(2)}_\mathrm{eff}$. Moreover, since this induced electric field is formed via the interference of the fundamental with its own second harmonic (Fig.~\ref{fig:overview}a), the direction of the field automatically switches with the correct periodicity for the QPM of second harmonic generation (SHG). While this technique provides a simplified method for fabricating a nonlinear grating for SHG, the low $\chi^{(2)}_\mathrm{eff}$ of SONGs in $\mathrm{SiO_2}$ limits the total conversion efficiency. More recently, the SONG concept has been reinvigorated through the observation of photo-induced SHG in nanophotonic waveguides made from silicon nitride ($\mathrm{Si_3N_4}$, hereafter SiN)~\cite{billat2017, porcel2017}, which offer strong spatial confinement of the light, massively scalable fabrication, and an improved $\chi^{(2)}_\mathrm{eff}$\cite{porcel2017}. However, all previous implementations of photo-induced SHG (in both $\mathrm{SiO_2}$ and SiN) were realized under conditions of strong group-velocity mismatch, limiting the phase-matching bandwidth, as well as the conversion efficiency for femtosecond pulses. 

Here we show that dispersion-engineered SiN photonic waveguides, which are currently enabling breakthroughs for ultrafast $\chi^{(3)}$ nonlinear optics \cite{carlson2017_calcium, carlson2017_efficient, ji2017, mayer2015, porcel2017_supercontinuum}, can also serve as a versatile platform for $\chi^{(2)}$ nonlinear optics with femtosecond pulses. Leveraging the high effective-index-contrast of air-top-clad SiN waveguides \cite{carlson2017_efficient}, we achieve group-velocity matching for SHG by engineering the waveguide cross section. Under these conditions, the formation of a SONG proceeds on the timescale of a few 10s of seconds and results in a QPM grating that supports the entire bandwidth of the femtosecond pulse with high conversion efficiency. Our model of photo-induced QPM is verified through the use of SHG microscopy to make the first direct observation of a SONG. Furthermore, we derive the equations that govern SONG formation for femtosecond pulses and confirm the crucial role of group-velocity matching. Finally, we demonstrate that a suitably prepared SiN waveguide can simultaneously generate light at the second-harmonic wavelength via $\chi^{(2)}$ and $\chi^{(3)}$ pathways, enabling $f$--$2f$ self-referencing of a laser frequency comb. This represents the first $f$--$2f$ stabilization of a frequency comb using a single amorphous material, and demonstrates how photonic waveguides can serve as an appealing platform for both $\chi^{(2)}$ and $\chi^{(3)}$ nonlinear optics with ultrafast sources.

\section{Results and Discussion}

\subsection{Photo-induced second harmonic generation}
\begin{figure}
	\includegraphics[width=\linewidth]{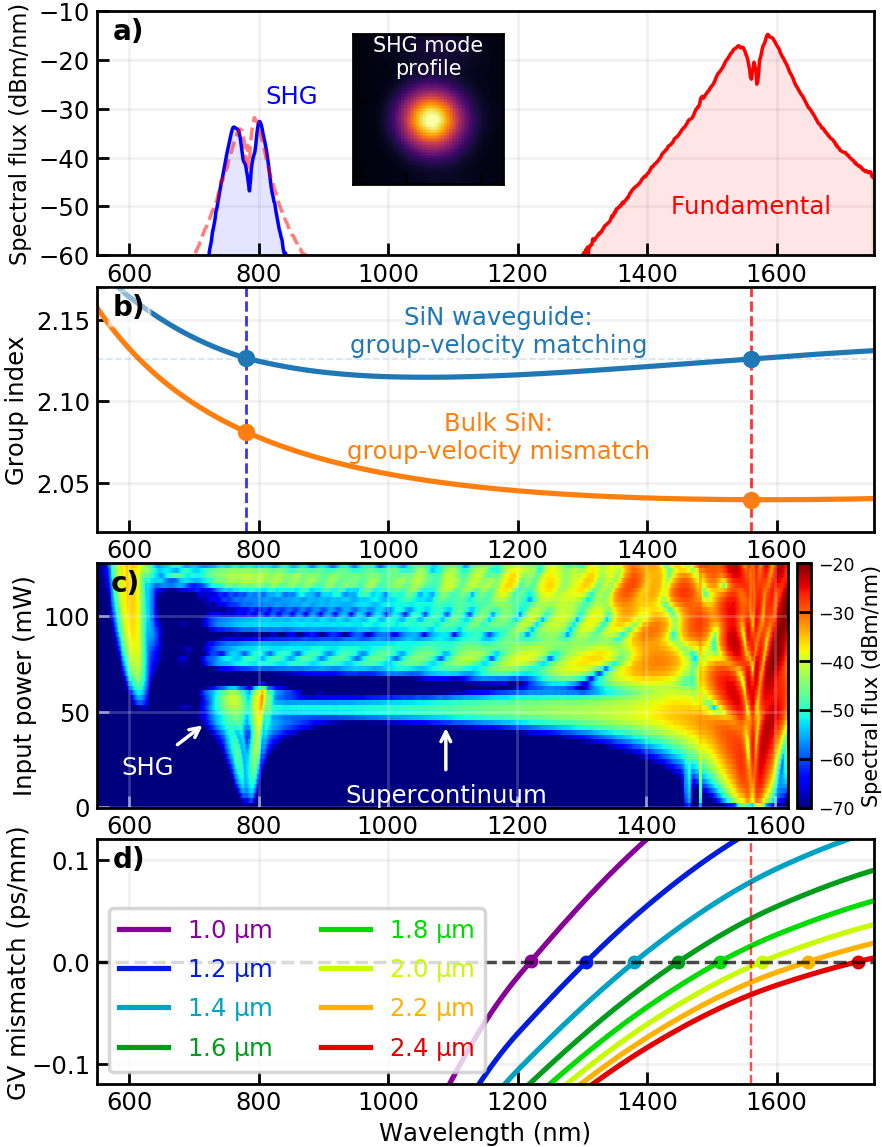}
	\caption{\label{fig:shg}
    \textbf{Second harmonic generation in amorphous silicon-nitride waveguides using femtosecond pulses.} a)~Experimentally, femtosecond pulses with a center wavelength of 1560~nm generate second-harmonic light near 780~nm. The dashed red line is the fundamental spectrum plotted at half the wavelength to demonstrate that the bandwidth is preserved. Inset: The output mode is in the fundamental quasi-transverse-electric ($\mathrm{TE_{00}}$) mode. b)~The group index of bulk silicon nitride monotonically decreases with wavelength, making group-velocity matching impossible. However, the group index of a 650$\times$1970-nm silicon-nitride waveguide provides perfect group velocity matching for 1560-nm SHG. c)~SHG is favored at relatively low input pulse intensities. At input powers over $\sim$60~mW, supercontinuum generation takes place and no SHG is seen. d)~ The fundamental-mode wavelength where group-velocity matching with the second-harmonic is achieved increases with increasing waveguide width.}
\end{figure}

We generate $\sim$200-fs pulses using a compact Er:fiber laser frequency comb with a center wavelength of 1560~nm and couple approximately 40~mW (400~pJ at 100~MHz repetition rate) into 12-mm-length, 650-nm-thickness silicon nitride waveguides \cite{carlson2017_calcium,carlson2017_efficient} that have an $\mathrm{SiO_2}$ bottom-cladding but air-cladding on the top and sides. After a build-up time of a few 10s of seconds (see Supplemental Fig.~\ref{fig:oscillations}), we observe broadband SHG with $\sim$1\% total conversion efficiency (Fig.~\ref{fig:shg}). Once the waveguide has been prepared, second harmonic light appears immediately upon turning the laser on. However, if the laser power is increased above approximately 60~mW, the supercontinuum generation process \cite{dudley2006,carlson2017_efficient} produces strong orange light (near 600~nm) and the SHG process is quickly quenched. However, decreasing the power below 40~mW allows the second harmonic to build up once again. This behavior indicates that the SHG results from the formation of a SONG through the coherent photogalvanic effect \cite{anderson1991, billat2017, porcel2017}.

While the exact microscopic mechanism that forms the static electric fields in SiN is not yet known, recent studies \cite{billat2017, porcel2017} have proposed that the separation of electrons and holes results in the static electric field. For the field to build-up in the waveguide between laser pulses -- and to remain permanently in the material -- electrons and holes must be stabilized at trap sites in the material. SiN is known to provide a high density of trap sites, which may hold electrons for more than 100 years \cite{tzeng2006}. Consequently, SiN has been used as a nonvolatile computer memory \cite{fujita1985}. Our SiN was deposited via low-pressure chemical vapor deposition (LPCVD), which has been shown to provide a higher density of trap sites than SiN deposited via plasma-enhanced CVD (PECVD) \cite{park1990}. 

Consistent with previous descriptions of SONG formation via trap-sites in SiN \cite{billat2017,porcel2017}, we find that the generation of visible-wavelength photons through the supercontinuum generation process can lead to the suppression of the SHG process, presumably by promoting trapped electrons to the conduction band and erasing the SONG. Similarly, a UV lamp focused on top of the waveguide can also lead to erasure (see Methods). Fortunately, the gratings appear to be unaffected by ambient laboratory light, even after several days of continuous illumination. 

\subsection{Group velocity matching}
In contrast to the previous studies of photo-induced SHG in SiN waveguides \cite{billat2017,porcel2017}, which observed SHG to higher-order modes in any waveguide, regardless of geometry, we only generate second harmonic across a small range of waveguide widths. This difference is a result of group-velocity matching, with the strongest SHG appearing for waveguide widths that allow the fundamental and the second harmonic to travel with the same group velocity. Importantly, the SHG from our group-velocity-matched waveguides has an output mode that does not contain any nodes (Fig.~\ref{fig:shg}a, inset), indicating that the second harmonic is generated in the fundamental quasi-transverse-electric ($\mathrm{TE_{00}}$) mode. 

For any waveguide SONG, the initial formation begins with a competition between the various spatial modes (fundamental and higher order) of the waveguide to determine which second-harmonic mode will be favored \cite{billat2017, porcel2017}. When pulses longer than a few picoseconds are used, the second harmonic is generated into the higher order mode that offers the lowest phase-mismatch. However, producing SHG into a higher order mode can be inconvenient, in part because it may reduce the conversion efficiency by decreasing the spatial overlap with the fundamental mode, but also because it complicates downstream applications, which generally prefer a beam profile without nodes.

If femtosecond pulses are instead used to form the SONG, the situation can be reversed, because the phase-mismatch must be considered across the entire bandwidth of the pump. Because phase matching to higher-order modes is intrinsically narrowband, there is strong group-velocity walkoff between the fundamental and higher-order modes. However, if a mode has perfect group-velocity matching, every wavelength in the pump spectrum has the same phase mismatch for SHG (neglecting higher-order dispersion), resulting in this mode being favored. Some of our waveguides achieve group-velocity matching to the fundamental mode, and consequently, our broad bandwidth femtosecond pulses preferentially form a grating with the correct periodicity to provide QPM for SHG to the fundamental mode. This is of critical importance, because, as we discuss later, the generation of SHG in the fundamental mode allows for good spatial overlap with light generated via supercontinuum generation.

The wavelength where group-velocity matching for SHG occurs can be tuned by changing the waveguide dimensions (Fig.~\ref{fig:shg}d). In our case, the 2000-nm waveguide offers the lowest group-velocity mismatch at the 1560-nm center wavelength of our laser. However, since we have a broadband pump, we find that we can generate SHG from waveguides with widths between 1900 and 2300~nm. As the width is increased, the peak of the second harmonic moves to longer wavelengths (Supplemental Fig.~\ref{fig:groupvelocity}), in agreement with the trend in the group-velocity matching (Fig.~\ref{fig:shg}d and Supplemental Fig.~\ref{fig:gv}).

Because the SONG can persist indefinitely, it can be used with other laser sources having a similar wavelength. We use this ability to directly map the phase-matching bandwidth of the gratings by using a few-milliwatt continuous-wave (CW) laser to generate SHG from each waveguide (Supplemental Fig.~\ref{fig:cw}). We observe broadband SHG, in some cases spanning 60~nm, by tuning the CW laser from 1520 to 1620~nm for each waveguide, confirming that group-velocity matching has been achieved. Additionally, as expected from Fig.~\ref{fig:shg}d, the peak conversion efficiency moves to longer wavelengths with increasing waveguide width. Thus, the group-velocity matching condition provides advantages for both femtosecond pulses and situations where tunability is required for CW frequency conversion.

\subsection{Microscopy of self-organized nonlinear gratings}

To confirm the presence of periodic $\chi^{(2)}$ gratings in our waveguides, we prepare the SONGs using the aforementioned procedure, irradiating each waveguide for approximately 10 minutes. We then transport the waveguides several kilometers to another laboratory, where we employ a SHG microscope (see Methods) to record the first direct images of a SONG. While the SHG experiment creates the periodic grating by propagating 1560-nm pulses along the waveguide (from left-to-right in Fig.~\ref{fig:overview}b,c), the microscope probes the sample from top to bottom with 870-nm pulses. The intensity of the emitted second-harmonic light is measured, and the laser is raster-scanned across the chip to form a two-dimensional image of the waveguide and the surrounding material. While most of the SiN material appears dark, the waveguides that exhibit strong SHG appear bright as a result of the strong induced $\chi^{(2)}_\mathrm{eff}$ (Fig.~\ref{fig:overview}b,c and Supplemental Fig.~\ref{fig:microscopy_images}). The observed periodicity of the SONG matches the period expected for QPM of 1560-nm SHG (Supplemental Fig.~\ref{fig:microscopy}). Moreover, we observe that the SONGs extend for \mbox{2--6 mm} (depending on the waveguide width), and that none of the SONGs exhibit any significant change in period along the length, confirming that group-velocity matching, and not variations in the grating period, is responsible for the broad-bandwidth SHG (Supplemental Fig.~\ref{fig:microscopy_length}). Our results are in general agreement with what has been observed when SONGs in $\mathrm{SiO_2}$ were visualized using electric-field-sensitive HF etching techniques \cite{margulis1995}. 

To further characterize the SONG, we record numerous SHG-microscopy images along the length of several waveguides and fit a sinusoidal function to the integrated SHG intensity (Supplemental Fig.~\ref{fig:microscopy_length}). We find that the intensity of the SONG follows a sigmoid function, where the SONG is not observable at the entrance facet, smoothly increases at some point along the waveguide length, and then reaches a constant value that is maintained until the exit facet. This behavior suggests that the SONG reaches a saturation condition, where the fundamental and SHG can no longer increase the intensity of the SONG. The origin of this saturation is not clear, though several possibilities exist. For example, the saturation may represent a condition where nearly all of the available trap sites in the material have been occupied. Alternatively, it may correspond to a situation where the static field is so large that it can cause charges to leave trap sites. Finally, it is possible that fundamental, second-harmonic, or third-harmonic photons promote trap-site electrons to the conduction band, and that the saturation effect occurs when an equilibrium is reached between the rate of promoting electrons to trap sites and the rate of promoting trap-site electrons to the conduction band. Future experiments that examine that nature of the saturation may provide insight into how the $\chi^{(2)}_\mathrm{eff}$ may be increased.

\subsection{Conversion efficiency}

For some waveguides, the SHG conversion exceeds 0.005~\%/Watt of peak power (Supplemental Fig.~\ref{fig:cw}), which is excellent conversion efficiency for a device that provides SHG across such a large bandwidth. Since we know the length of the grating from the microscopy experiments, we can use Eqs.~9 and 10 from Ref.~\citenum{porcel2017} to calculate the $\chi^{(2)}_\mathrm{eff}$ to be 0.5~pm/V for the 2100~nm waveguide. 

It is likely that the lengths of the SONGs that we observe are limited by the broad bandwidth of our laser pulses, and that much longer gratings could be produced if narrower-bandwidth pulses propagated through longer waveguides. A longer grating would provide a narrower phase-matching bandwidth, but greatly enhanced conversion efficiency.

\begin{figure}
	\includegraphics[width=\linewidth]{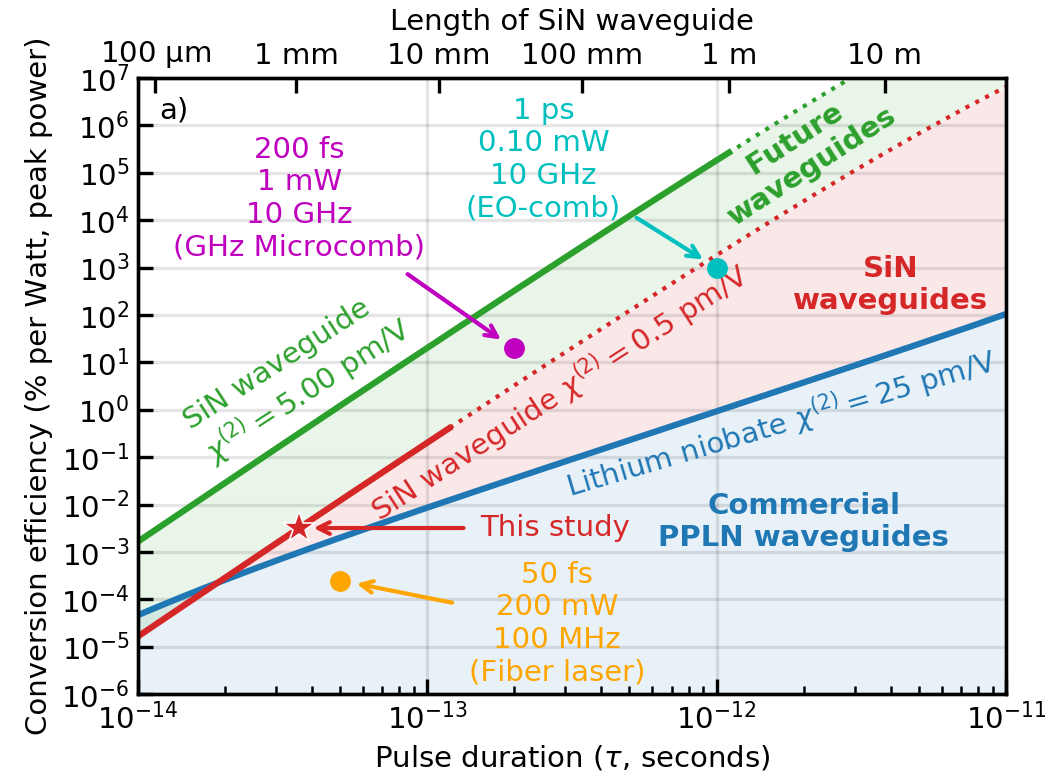}
	\includegraphics[width=\linewidth]{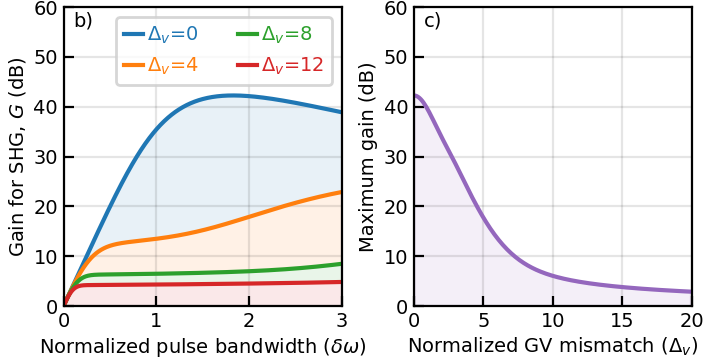}
	\caption{\label{fig:theory} \textbf{Analytical estimate of SHG with SiN waveguides.} a) Using the $\chi^{(2)}_\mathrm{eff}$ obtained in this study and the calculated dispersion of a 1953$\times$649-nm SiN waveguide (which offers group-velocity matching at 1560-nm), we can plot the conversion efficiency of SiN waveguides as a function of the pulse duration. The red line indicates the performance of the SiN waveguides as demonstrated in this work, while the green line represents the potential performance if an increase in material $\chi^{(3)}$ or static electric field could enable $\chi^{(2)}_\mathrm{eff}$ to be increased to 4.71 pm/V. In comparison to a commercial SHG waveguide made from periodically poled lithium niobate (PPLN, blue), the SiN waveguides provide a $\tau^4$ scaling of the conversion efficiency rather than a $\tau^2$ scaling of a group-velocity-mismatched material. b) The gain for SHG as a function of the input pulse bandwidth $\delta \omega$ for various values of the normalized group-velocity mismatch $\Delta_v$ demonstrates that higher gain is provided by smaller values of group-velocity mismatch. In general, when the group-velocity mismatch is higher, a smaller pulse bandwidth is required to reach higher gain. However, as the pulse bandwidth is increased to very large values, the gain increases once again, because now light is present at the frequency where group-velocity matching is achieved. c) For photo-induced SHG with femtosecond pulses, the highest conversion efficiency will be realized through perfect group-velocity matching. }
\end{figure}

Using our calculated $\chi^{(2)}_\mathrm{eff}$, we can make a comparison to a periodically poled lithium niobate (PPLN) waveguide with $\chi^{(2)}=50$~pm/V and an effective mode area of \SI{100}{\micro\meter}$^2$. To our knowledge, such PPLN waveguides provide the highest conversion efficiency for any commercially available nonlinear optical device. Fig.~\ref{fig:theory}a shows the maximum conversion efficiency achievable, while preserving the full bandwidth of the pulse. As the pulse duration becomes longer, a SiN waveguide should outperform a PPLN waveguide, since the group-velocity matching allows a longer grating to be supported. Given the optimal length for both devices, a SiN waveguide should outperform a PPLN waveguide for pulse durations longer than about 30 fs. If the $\chi^{(2)}_\mathrm{eff}$ could be further increased, SONGS based on SiN could enable SHG with even lower pulse energies, enabling frequency conversion of un-amplified electro-optic \cite{carlson2018} and microresonator-based \cite{lamb2018} frequency combs. The dots in Fig.~\ref{fig:theory}a represent the conversion efficiency required to achieve 10\% conversion for various input pulse energies and durations.

\subsection{Analytical description of grating formation}

While several models exist for photo-induced grating formation in the case of CW light \cite{anderson1991, zeldovich1989, baranova1990}, to our knowledge, no theoretical treatment has been attempted for femtosecond pulses. Thus, we explore the photo-induced grating formation process analytically, following the procedure of Anderson et al. \cite{anderson1991}, and present the complete analysis in Supplemental Materials Section \ref{sec:supp_theory}. By assuming that we can model the total photo-induced grating as a coherent sum of contributions from all of the individual frequencies, we find that the second-harmonic pulse energy increases from the noise level at the input end $U_2(0)$ to the output value of $U_{2,\mathrm{out}}=U_2(0)\exp(G)$ with the total logarithmic gain for SHG given by

\begin{equation}
\label{eq:gaint}
G(\delta \omega, \Delta_v) \sim G_0 \delta\omega \int_0^1 \frac{e^\frac{-(\Delta_v\delta\omega z')^2}{2(1+\delta\omega^4z'^2)}}{\sqrt{1+\delta \omega^4z'^2}} dz',
\end{equation}

where $\delta \omega = \Delta \omega \sqrt{\delta\beta L}$ is the normalized pulse bandwidth,  $\Delta \omega$ is the actual pulse bandwidth, $L$ is the length of the waveguide, and $\delta \beta = \beta_{2\omega_0} - \beta_{\omega_0}$ is the difference in group-velocity dispersion (GVD) at the fundamental and second-harmonic frequencies. $\Delta_v=\delta v_g^{-1} \sqrt{L/\delta\beta}$ is the normalized group-velocity mismatch, and $\delta v_g^{-1} = v_{g, 2\omega_0}^{-1} - v_{g,\omega_0}^{-1}$ is the (inverse) group-velocity mismatch between the fundamental and the SHG. The coefficient $G_0$ incorporates several material characteristics, including the effective coherent photoinjection coefficient, third-order nonlinear susceptibility, momentum scattering and recombination times, trapping cross-section, and the effective cross-section of the waveguide, which makes theoretical estimate of $G_0$ difficult. However, comparing our results with those of Ref.~\citenum{anderson1991}, it is not unreasonable to assume that our gain is similar, i.e. on the order of 40~dB. 

Eq.~\ref{eq:gaint} indicates that better group-velocity matching (lower $\Delta_v$) allows for larger maximum gain. Moreover, better group-velocity matching allows for SHG with shorter pulses, explaining why previous studies observed SHG with picosecond pulses, and how the group-velocity matching enabled by nanophotonic waveguides allows for SHG with femtosecond pulses (Fig.~\ref{fig:theory}b,c).

\subsection{Simplified frequency comb stabilization}

\begin{figure}
    \includegraphics[width=\linewidth]{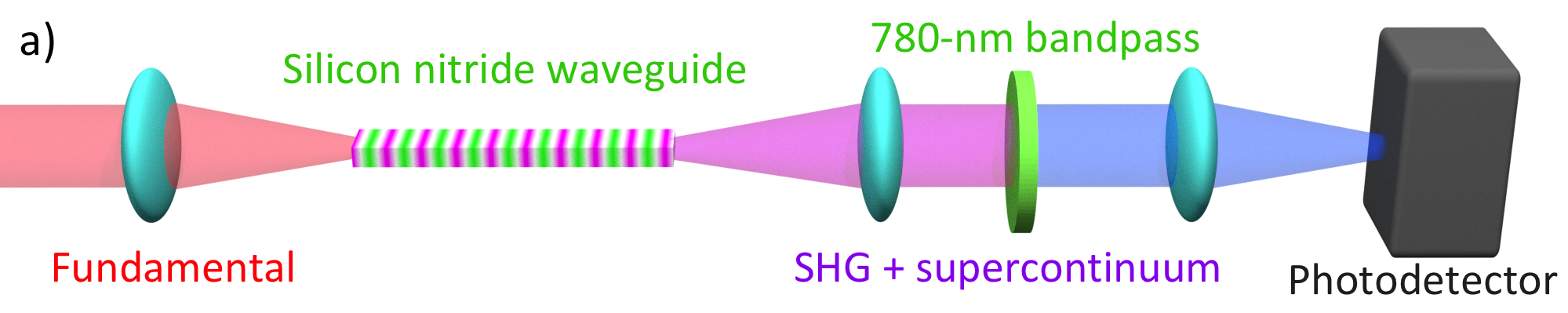}
    \includegraphics[width=\linewidth]{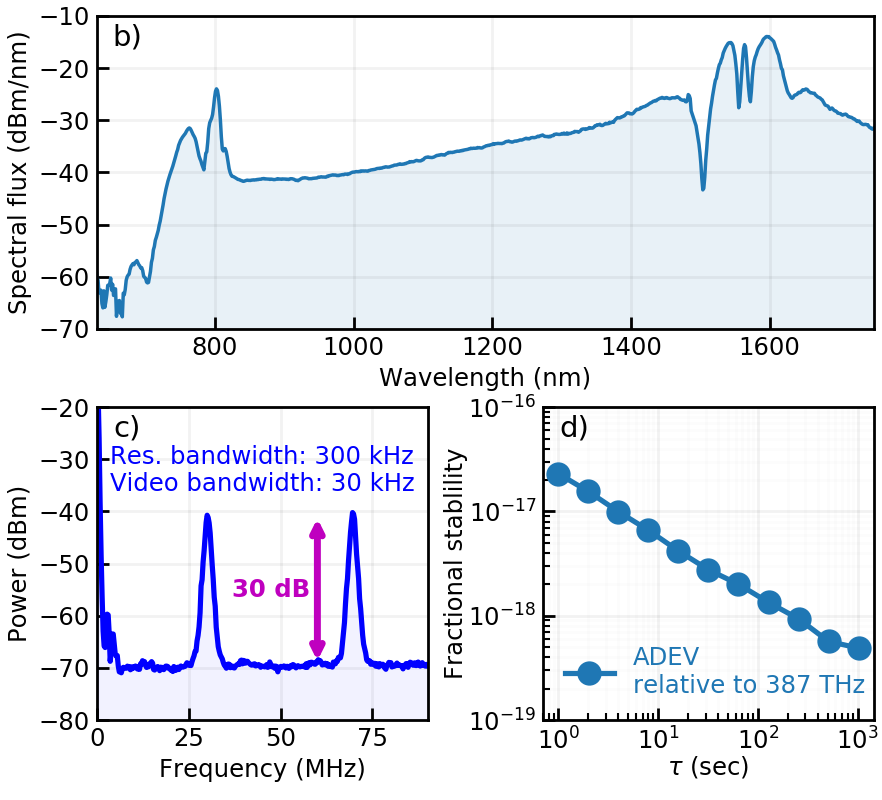}
    \caption{\label{fig:f2f} \textbf{Frequency comb stabilization through one-step $f$-$2f$ self-referencing.} a) In contrast to a typical $f$-$2f$ self-referencing experiment, which requires separate SHG and supercontinuum-generation media, a single SiN waveguide provides both $f$ and $2f$ light, which can be selected using a simple filter and detected with a silicon photodetector. b) The input power is set at a level that provides both supercontinuum and second-harmonic light near 800 nm. c) The carrier--envelope offset frequency ($f_\mathrm{ceo}$) is detected with 30 dB signal-to-noise ratio, which is sufficient for stabilizing the frequency comb without phase slips. c) The Allan deviation of the locked $f_\mathrm{ceo}$ (recorded with a separate photodetector and frequency counter) confirms that $f_\mathrm{ceo}$ has been stabilized to a level suitable for precision measurements.}
\end{figure}

A particular advantage of SiN nanophotonic waveguides is that they exhibit simultaneous supercontinuum generation ($\chi^{(3)}$) and SHG ($\chi^{(2)}$), which allows for \mbox{$f$-$2f$} self-referencing of frequency combs. In previous experiments\cite{hickstein2017}, waveguides made from aluminum nitride, a material with bulk $\chi^{(3)}$ and $\chi^{(2)}$ provided both SHG and supercontinuum generation and allowed for a simplified scheme for self-referencing. However, nJ pulse energies were required, since SHG to the fundamental mode was strongly phase mismatched and was consequently very dim. In contrast, in the case of SiN SONGs, the SHG is fully quasi-phase-matched and the conversion efficiency is much higher, which allows lower pulse energies to be used. For production of both SHG and supercontinuum at 780-nm, the pulse energy must be set appropriately. If it is too low, then insufficient supercontinuum is generated, and if it is too high, then the SHG intensity is diminished. Fortunately, a regime exists where both supercontinuum and SHG are generated with sufficient intensity to detect the carrier--envelope-offset frequency ($f_\mathrm{ceo}$) with $\sim$30 dB signal-to-noise ratio (SNR, Fig.~\ref{fig:f2f}). We stabilize $f_\mathrm{ceo}$ by feeding back to the laser current and verify that performance equivalent to a traditional $f$-$2f$ interferometer can be achieved by counting the in-loop beat note with a frequency counter \cite{carlson2017_efficient}.

\section{Outlook}
We have performed a proof-of-principle demonstration, showing that nanophotonic waveguides can provide broadband SHG for femtosecond pulses. Looking to the future, there are several open questions and possible avenues for improvement. For instance, it should be possible to achieve much higher conversion efficiencies than demonstrated here by increasing either $\chi^{(2)}_\mathrm{eff}$ or the length of the waveguide. The $\chi^{(2)}_\mathrm{eff}$ could be optimized via several methods including increasing the material $\chi^{(3)}$, engineering the material to support higher electric fields, or finding a material that has a stronger $\chi^{(2)}_\mathrm{eff}$ response via slight re-arrangement of the atomic positions. As an example of how a high $\chi^{(3)}$ can result in a high $\chi^{(2)}_\mathrm{eff}$, a recent study \cite{timurdogan2017} has shown that silicon, a material with a strong $\chi^{(3)}$, can be biased with electrodes to enable a $\chi^{(2)}_\mathrm{eff}$ of 41~pm/V.

Increasing the length of the waveguide first requires that the waveguide loss is low. Fortunately, recent efforts have realized SiN waveguides with losses on the order of 1~dB/meter \cite{ji2017}, and suggest that the losses may be decreased orders-of-magnitude further before reaching the material limit. Second, increasing the length of the waveguide while preserving the bandwidth of the SHG requires that the waveguide offers low phase mismatch over the entire bandwidth of the input pulse. This amounts to a waveguide that exhibits low and flat GVD over a broad spectral region. Fortunately, nanophotonic waveguides provide exceptional control over GVD, and several studies \cite{zhu2012, zhang2010} have shown that ``slot waveguide'' geometries can be used to provide ultra-flat GVD profiles. If both $\chi^{(2)}_\mathrm{eff}$ and the waveguide length can be significantly increased, nanophotonic waveguides could enable nonlinear optics with ultra-low pulse energies, which would have numerous applications, ranging from quantum computing \cite{weston2016} to microresonator frequency combs \cite{lamb2018}.

\section{Conclusion}

We have shown that nanophotonic waveguides can provide quasi-phase-matching for nonlinear processes via self-organized nonlinear grating formation. Unlike previous observations of such gratings, we form the gratings using femtosecond pulses, and observe broadband second harmonic generation to the fundamental mode of the waveguide. These nanophotonic waveguides provide both quasi-phase-matching and group-velocity matching for second harmonic generation, opening the door for dispersion-engineered $\chi^{(2)}$ ultrafast nonlinear optics in amorphous waveguides. Using second-harmonic-generation microscopy, we record the first direct images of self-organized nonlinear gratings. Additionally, by analytically deriving the equations that govern the formation of the nonlinear gratings, we quantify the role of group-velocity, and group-velocity-dispersion mismatch in the self-organization process. Finally, we demonstrate the utility of such nanophotonic waveguides by self-referencing a laser frequency comb with a single waveguide. In the future, longer and more nonlinear waveguides will enable high-efficiency nonlinear optics in situations previously considered impossible.

\begin{acknowledgments}
We acknowledge Gregory Moille, and the NIST Boulder Editorial Review Board for providing helpful feedback on this manuscript. We thank Kevin Dorney and Jennifer Ellis, Henry Kapteyn, and Margaret Murnane for the timely loan of a polarizer. We acknowledge Norman Sanford for insightful discussions. This work is supported by AFOSR under award number FA9550-16-1-0016, DARPA (DODOS and ACES programs), NIST, and NRC. This work is a contribution of the U.S. government and is not subject to copyright in the U.S.A.  
\end{acknowledgments}

\section{Methods}

\subsection{Waveguide design}

The waveguide fabrication took place at NIST in Gaithersburg, Maryland. The waveguides are composed of $\mathrm{Si_3N_4}$ deposited via low-pressure-chemical-vapor-deposition (LPCVD) on top of a \SI{3}{\micro\meter} thermal $\mathrm{SiO_2}$ layer, which is supported by a silicon substrate. Two different chips of waveguides were investigated for SHG. One sample consists of 650-nm thick ``air-clad'' waveguides without side or top cladding. The other sample consists of 700-nm thick waveguides that are ``air-clad'' in the central region, but have an $\mathrm{SiO_2}$ top and side cladding in the first and last 1~mm of the waveguide length to improve the coupling efficiency. Details regarding the 600-nm thickness waveguides are provided in Ref.~\citenum{carlson2017_calcium}, while details of the 700-nm thickness waveguides are provided in Ref.~\citenum{carlson2017_efficient}. The waveguide modes (and their effective indices) are calculated using a vector finite-difference modesolver \cite{fallahkhair2008, bolla2017}, using published refractive indices for $\mathrm{Si_3N_4}$ \cite{luke2015} and $\mathrm{SiO_2}$ \cite{malitson1965}.

\subsection{Second harmonic generation}

Second harmonic generation experiments were completed at NIST in Boulder, Colorado. We generate second harmonic by coupling 1560-nm light from a compact 100~MHz Er-fiber frequency comb \cite{sinclair2015} into each SiN waveguide. The power is adjusted using a computer-controlled rotation-mount containing a half-waveplate, which is placed before a polarizer. The polarization is set to horizontal (along the long dimension of the rectangular waveguide) and excites the lowest order quasi-transverse-electric ($\mathrm{TE_{00}}$) mode of the waveguide. 

Interestingly, in some cases, we see strong temporal oscillations in the intensity of the second harmonic. Similar oscillations were also observed in other studies of silicon nitride waveguides\cite{billat2017, porcel2017}. In our waveguides, the oscillations typically become faster when the laser intensity is increased (Supplemental Fig.~\ref{fig:oscillations}). Additionally, they are nearly absent for the waveguides that exhibit the best group-velocity matching, suggesting that they are related the the group-velocity matching conditions. We suspect that these oscillations result from slight changes in the period of the grating during the formation process. For example, an initial grating might form with a period that phase-matches SHG at the peak of the pump spectrum. Over time, the grating may slightly change period to provide phase matching for the wavelength that experiences the best group-velocity matching in the waveguide.

\subsection{Frequency comb stabilization}
We used the SiN waveguides to provide a simplified, low-power method to stabilize our Er-fiber frequency comb. By setting the power level to $\sim$40~mW, we generate light near 780~nm via both SHG (a $\chi^{(2)}$ process) and supercontinuum generation (a $\chi^{(3)}$ process). The interference of these two pathways allows $f_\mathrm{ceo}$ to be detected simply by detecting the light near 780~nm. This approach is in contrast to conventional $f$-$2f$ self-referencing, which separately generates the supercontinuum light in one material and the second harmonic light in another material \cite{carlson2017_efficient}.

The light emitted by the waveguide was collimated with a microscope objective with a numerical aperture (NA) of 0.85 (Newport M60x) and filtered with a 780-nm bandpass filter (Thorlabs FB-780-10). A beamsplitter directed the light to two separate silicon avalanche photodiodes (APD, Thorlabs APD430A and APD210). The electrical signal from the first photodiode was connected to a Red Pitaya field-programmable-gate-array (FPGA) board running the ``Frequency comb digital-phase-locked-loop'' firmware \cite{sinclair2015, tourigny2018}, which was used to feed back to the oscillator pump-diode current in order to stabilize $f_\mathrm{ceo}$. The electrical signal for the second APD was connected to a $\Lambda$-type frequency counter, which provided an ``electrical out-of-loop'' confirmation of the $f_\mathrm{ceo}$ stabilization. 

\subsection{Second-harmonic-generation microscopy}
Second harmonic generation microscopy experiments were completed at the Smalyukh Lab at the University of Colorado. A Ti:sapphire laser (Coherent Chameleon Ultra II) was used to generate 140-fs pulses with a central wavelength of 870~nm at an 80-MHz repetition-rate. The laser was attenuated so that, at the sample, the average power was 6 mW, corresponding to a pulse energy of 75~pJ. The beam enters an inverted microscope system (Olympus IX81) and is focused onto the sample using an Olympus UPlanFLN 40x objective with NA=0.75, providing a peak intensity of approximately $4.7\times 10^{10}\, \mathrm{W/cm^2}$, assuming a diffraction-limited spot size. The focused spot was raster-scanned across the sample using a galvo mirror system (Olympus Fluorview FV300). The reflected second-harmonic light propagates back into the galvo-mirror system where it is collected with a photomultiplier tube. Images were collected at a rate of $\sim$32,000~pixels/second and a size of $512\times 512$~pixels, for an image acquisition rate of $\sim$9~s per image. The images are averaged using a 10-frame Kalman filter, for a total acquisition time of $\sim$90~s per averaged image. 

Generally, the SHG microscopy experiment did not significantly alter the self-organized gratings, indicating that it may be possible to record such images \textit{in situ}, while the grating is forming. However, in several cases, when the laser was left to raster over a small region for many minutes, we noticed a slow decrease in the intensity of the SHG being emitted from the waveguide, indicating that the grating was being erased. It is not clear if the erasure is caused by the high peak intensity of the femtosecond pulses or simply the average power. Additionally, during an initial microscopy attempt, we aligned the microscope by illuminating the chip with a 100~W mercury-vapor lamp (Olympus U-LH100HG), which emits high power visible and UV light. However we found that the focused light from this lamp appeared to rapidly erase the self-organized gratings. However, we found that illumination with a lower-power halogen lamp (Olympus LG-PS2-5), which produces much lower levels of UV light, allowed the microscope to be aligned without affecting the observed SHG, indicating that irradiation with high-energy photons is an effective method for grating erasure.

\subsection{Estimation of the effective quadratic nonlinearity}

Using a continuous wave (CW) laser, we can generate second harmonic from the waveguides that have been suitably prepared by the femtosecond-pulsed laser. Because the CW laser doesn't spectrally broaden as it propagates along the waveguide, it allows an accurate measurement of the SHG conversion efficiency. Additionally, since the SHG microscopy experiment provides estimates of the length of the grating in each waveguide, we can estimate the effective quadratic nonlinearity of our SiN waveguides. According to Eq.~5.37 of Weiner \cite{weiner2009} and Eq.~3.19 of Suhara and Fujimura \cite{suhara2003}, we can write the conversion efficiency for SHG as 
\begin{equation}
\label{eq:efficiency}
\frac{\eta}{P_\omega} = \frac{L^2 d_\mathrm{eff}^2}{S_\mathrm{eff}} \left(\frac{2 \omega_0^2}{\varepsilon_0 c^3 n^3}\right),
\end{equation}
where $\eta$ is the conversion efficiency, $P_\omega$ is the peak power of the fundamental, $L$ is the medium length, $\varepsilon_0$ is the permittivity of free space, $\omega_0$ is the frequency of the fundamental, $c$ is the speed of light, $n$ is the index of the fundamental, $S_\mathrm{eff}$ is the effective mode area, and the nonlinearity $d_\mathrm{eff} = 1/2 \chi^{(2)}_\mathrm{eff}$.

We can see from Eq.~\ref{eq:efficiency} that increasing the medium length is equally important to increasing $d_\mathrm{eff}$. Re-arranging Eq.~\ref{eq:efficiency} to solve for $d_\mathrm{eff}$ yields

\begin{equation}
\label{eq:deff}
d_\mathrm{eff} = \sqrt{\left( \frac{\eta}{P_\mathrm{\omega}} \right) \left( \frac{S_\mathrm{eff}}{L^2}\right) \left(\frac{\varepsilon_0 c^3 n^3}{2\omega_0^2}  \right) }.
\end{equation}
Using a numerical vector finite difference modesolver \cite{fallahkhair2008,bolla2017}, we calculate that the effective area for 1560-nm SHG in a 650$\times$2100-nm waveguide is \SI{1.20}{\micro\meter}$^2$. Using the observed grating length (1.05~mm) and the observed conversion efficiency (0.0025~\%/W), we can use Eq.~\ref{eq:deff} to calculate $d_\mathrm{eff} = 0.25$~pm/V ($\chi^{(2)}$ = 0.5~pm/V). This is in rough agreement with the value of $\chi^{(2)}$ = 0.3~pm/V reported by Billat, et al. \cite{billat2017}.

\section{Disclaimer}
Certain commercial equipment, instruments, or materials are identified here in order to specify the experimental procedure adequately. Such identification is not intended to imply recommendation or endorsement by the National Institute of Standards and Technology, nor is it intended to imply that the materials or equipment identified are necessarily the best available for the purpose. This work is a contribution of the United States government and is not subject to copyright in the United States of America.

\section{References}
\bibliography{shg}

\clearpage
\onecolumngrid
\begin{center}
\textbf{\large Supplemental Materials}
\end{center}
\twocolumngrid

\setcounter{equation}{0}
\setcounter{figure}{0}
\setcounter{table}{0}
\setcounter{section}{0}
\makeatletter
\renewcommand{\thesection}{S\arabic{section}}
\renewcommand{\theequation}{S\arabic{equation}}
\renewcommand{\thefigure}{S\arabic{figure}}

\section{Theory of photo-induced grating formation with femtosecond pulses}
\label{sec:supp_theory}

\begin{figure}
	\includegraphics[width=\linewidth]{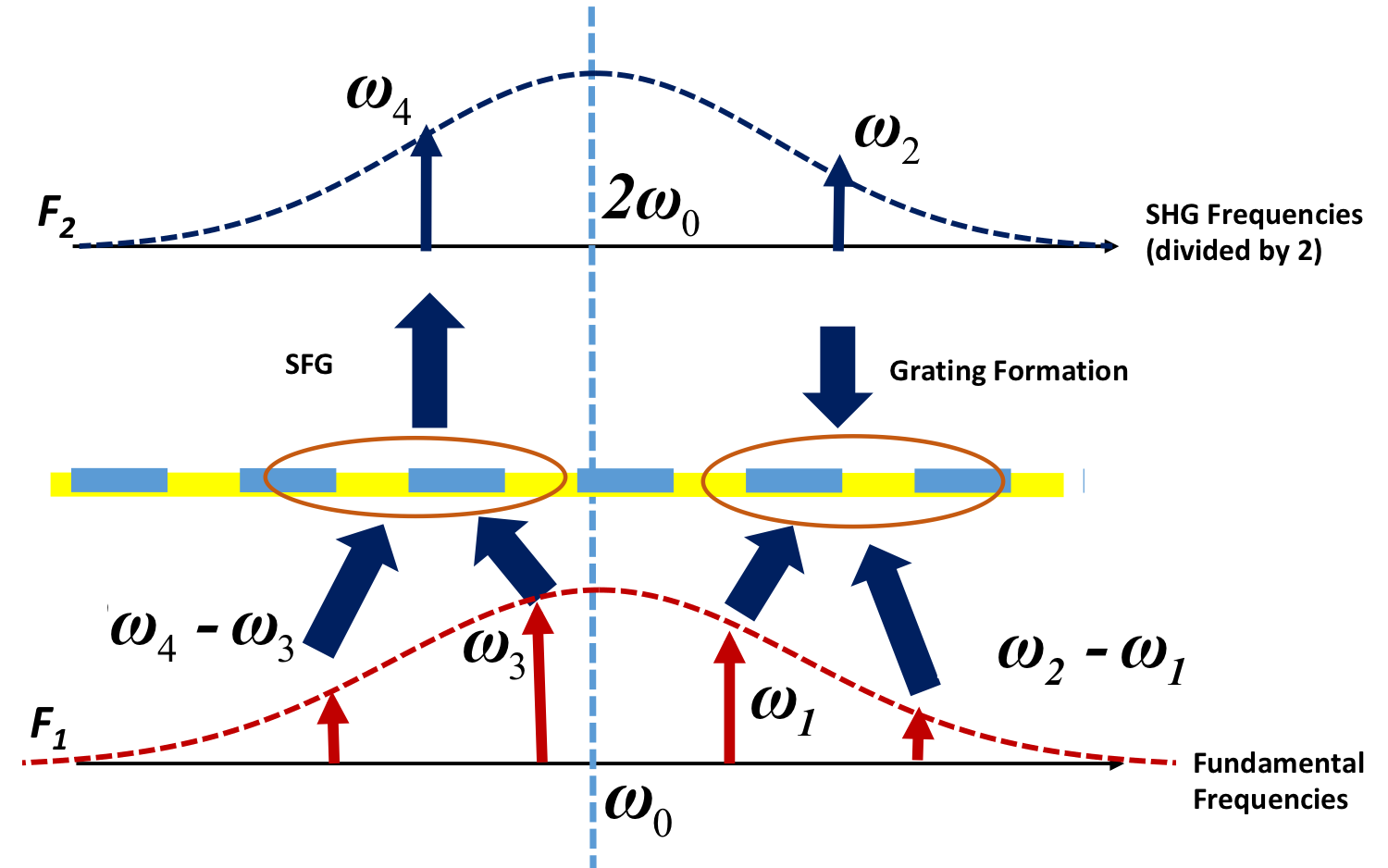}
	\caption{\label{fig:theory_sketch} \textbf{The process of self-phase matched photoinduced SHG with femtosecond (broadband) pulses.} The interference of two photon absorption of fundamental photons at frequencies $\omega_1$ and $(\omega_2 - \omega_1)$ with a single photon absorption at frequency $\omega_2$ produces quantum interference current and the grating of effective $\chi^{(2)}$. This grating serves to achieve quasi-phase-matching (QPM) for sum frequency generation with fundamental frequencies $\omega_3$ and $\omega_4-\omega_3$ producing  frequency component near SHG frequency $2\omega_0$. }
\end{figure}

The grating formation is driven by the interaction of the fundamental pulse $E_1(t)$, centered at carrier frequency $\omega_0$ (wavevector $k_{\omega_0}$) and the second-harmonic (SH) pulse $E_2(t)$, centered at twice the carrier frequency $2\omega_0$ (wavevector $k_{2\omega_0}$)(Fig.~\ref{fig:theory_sketch}). Both pulses can be represented in the frequency domain as a superposition of monochromatic waves:
\begin{align}
\begin{split}
E_1(t) &= e^{i(k_{\omega_0} z - \omega_0 t)} \int F_1(\omega) e^{i(k_1z-\omega_1t)}d\omega_1\\
E_2(t) &= e^{i(k_{2\omega_0} z - 2\omega_0 t)} \int F_2(\omega) e^{i(k_2z-\omega_2t)}d\omega_2
\end{split}
\end{align}
where $\omega_{1,2}$ and $k_{1,2}$ are the frequencies and wavevectors relative to the carriers of the fundamental and the SH respectively and $F_1$ and $F_2$ are the Fourier transforms of each field.

Three-wave interference of one photon near the SH carrier frequency and two photons near the fundamental carrier frequency creates a coherent, directional photocurrent $J_\mathrm{coh}$ proportional to the vector potentials of these three waves,
\begin{align}
\begin{split}
\label{eq:current}
J_\mathrm{coh} &\propto A_1(\omega_1)A_1(\omega_2 - \omega_1) A_2^*(\omega_2)e^{i\Delta k(\omega_1, \omega_2)}\\
& \propto iF_1(\omega_1)F_1(\omega_2 - \omega_1)F_2^*(\omega_2)e^{i\Delta k(\omega_1, \omega_2)}.
\end{split}
\end{align}
 The fact that $J_\mathrm{coh}$ depends on the vector potentials and not the electric fields \cite{khurgin1995} produces a 90-degree phase shift in Eq.~\ref{eq:current}. This 90-degree phase shift is crucial for enabling constructive build-up of the static electric field \cite{anderson1991}.
 The phase mismatch in Eq.~\ref{eq:current} is given by
\begin{align}
\begin{split}
\label{eq:delta_k}
\Delta k (\omega_1, \omega_2) = \Delta k_0 &+ \delta v_g^{-1} \omega_2 + \delta \beta \omega_2^2\\
 	&+2\beta_{\omega_0} \omega_1(\omega_2 - \omega_1),
\end{split}
\end{align}
where $\Delta k_0 = k_{2\omega_0} - 2k_{\omega_0}$ is the mean momentum mismatch, $\delta v_g^{-1} = v_{g, 2\omega_0}^{-1} - v_{g,\omega_0}^{-1}$ is the (inverse) group velocity mismatch between the fundamental and the SH, and $\delta \beta = \beta_{2\omega_0} - \beta_{\omega_0}$ is the difference in group-velocity dispersion (GVD) at the fundamental and SH frequencies.

For the silicon nitride waveguides, dispersion engineering is used to reduce $\delta v_g^{-1}$, and this typically causes the GVD at the SH frequency to have similar magnitude, but the opposite sign to the GVD at the pump, hence $\delta \beta \approx 2 \beta_{\omega_0}$. At the same time, $\omega_1(\omega_2 - \omega_1) \leq \omega_2^2/4$, which means that the last term on the right side of Eq.~\ref{eq:delta_k} is less than 10\% of the third one, and we neglect it in the following analysis and consider the phase-mismatch to be
\begin{equation}
\label{eq:delta_k2}
\Delta k(\omega_2) = \Delta k_0 + \delta v_g^{-1} \omega_2 + \delta \beta \omega_2^2.
\end{equation}

According to the model developed in Anderson et al. \cite{anderson1991}, the photo-excited carriers get trapped and give rise to a DC field with a sinusoidal spatial pattern with a period of $\Delta k$
\begin{align}
\begin{split}
E_\mathrm{DC} (\omega_2, z) \sim &\, iCF_2^*(\omega_2)e^{i\Delta k(\omega_2)z} \\
	&\times \int F_1(\omega_1)F_1(\omega_2-\omega_1)d\omega_1 + \mathrm{c.c.},
\end{split}
\end{align}
where $\mathrm{c.c.}$ denotes the complex conjugate. $C$ is a phenomenological coefficient that incorporates current injection efficiency as well as scattering and recombination times to describe the saturation effect due to excitation of photocarriers that do not exhibit the spatial interference pattern. According to the Wiener-Khinchin theorem, the auto-correlation of the fundamental spectrum is equal to

\begin{align}
\begin{split}
\label{eq:wk}
\int F_1(\omega_1) F_2(\omega_2-\omega_1) d\omega_1 &= \frac{1}{2 \pi}\int \left|E_1(t)\right|^2 e^{-i\omega_2 t} dt \\
	&\equiv F_1^2(\omega_2),
\end{split}
\end{align}
 where $F_1^2(\omega_2)$ is the Fourier transform of the square of the pulse and not the square of the absolute value of the Fourier transform ($\left|F_1(\omega_1)\right|^2$). Thus, the total DC field can be written as a superposition of different gratings,

\begin{equation}
\label{eq:dc}
E_\mathrm{DC} \sim iC\int F_1^2 (\omega_2) F_2^* (\omega_2) e^{i\Delta k(\omega_2)z} d\omega_2 + \mathrm{c.c.}
\end{equation}

This DC grating will induce the effective second order susceptibility $\chi^{(2)}(2\omega;\omega; \omega; z) \sim E_\mathrm{DC}(z) \chi^{(3)}(2\omega;\omega; \omega, 0)$, allowing us to write the equation for the growth of the SH frequency component $F_2(\omega_4)$ that arises as a result of sum frequency generation of different Fourier components of the fundamental pulse at frequencies $\omega_3$ and $\omega_4-\omega_3$
\begin{align}
\begin{split}
\label{eq:df2}
\frac{dF_2(\omega_4)}{dz} \sim & -i\chi^{(3)} E_\mathrm{DC}(z) \\
& \times \int F_1(\omega_3)F_1(\omega_4-\omega_3) e^{-i\Delta k(\omega_4)z}d\omega_3
\end{split}
\end{align}
where, in accordance with Eq.~\ref{eq:delta_k2},
\begin{equation}
\Delta k(\omega_4) = \Delta k_0 + \delta v_g^{-1} \omega_4 + \delta \beta \omega_4^2.
\end{equation}
Substituting Eq.~\ref{eq:dc} into Eq.~\ref{eq:df2} and using Eq.~\ref{eq:wk}, we obtain
\begin{align}
\begin{split}
\label{eq:f2}
\frac{dF_2(\omega_4)}{dz} \sim CF_1^2(\omega_4) \int F_1^2(\omega_2)F_2^*(\omega_2)e^{i\Delta k(\omega2, \omega_4)z}d\omega_2,
\end{split}
\end{align}
where
\begin{equation}
\Delta k(\omega_2, \omega_4) = \delta v_g^{-1} (\omega_2 - \omega_4) + \delta \beta (\omega_2^2 - \omega_4^2)
\end{equation}
and coefficient $C$ now also includes $\chi^{(3)}$. Next, we multiply both sides of Eq.~\ref{eq:f2} by $F_2^*(\omega_4)$ and integrate over $\omega_4$, taking into account the fact that the energy of the SH pulse is given by $U_2 = \int \left|F_2(\omega)\right|^2d\omega$:
\begin{align}
\begin{split}
\label{eq:du}
\frac{dU_2}{dz} &\sim C \int \int F_1^2 (\omega_2) F_1^{2*}(\omega_3)F_2^*(\omega_2)F_2(\omega_4) \\
		&\;\;\;\; \times e^{i\Delta k(\omega_2, \omega_4)z}d\omega_2 d\omega_4\\
	&= C \left| \int F_1^2(\omega_2)F_2^*(\omega_2)e^{i(\delta v_g^{-1}\omega_2+\delta\beta\omega_2^2)z}d\omega_2 \right|^2
\end{split}
\end{align}

Let us assume that the fundamental pulse shape is a Fourier-limited Gaussian pulse of energy $U_1$ and duration $\tau$ and that the SH pulse electric field follows the square of the fundamental pulse according to
\begin{align}
\begin{split}
\label{eq:gauss_time}
E_1^2(t) &= \frac{U_1}{\sqrt{2\pi} \tau} e^{-t^2/2\tau^2}\\
E_2(t) &\propto E_1^2(t) = \left( \frac{U_2}{\pi^{1/2} \tau}\right)^{1/2} e^{-t^2/2\tau^2}.
\end{split}
\end{align}
Accordingly, 
\begin{align}
\begin{split}
\label{eq:gauss_freq}
F_1^2(\omega) &= \frac{U_1}{2\pi} e^{-\omega^2\tau^2/2}\\
F_2(\omega) &= \left( 2 \pi^{1/2} \tau U_2\right)^{1/2} e^{-\omega^2\tau^2/2}
\end{split}
\end{align}
substituting Eq~\ref{eq:gauss_freq} into Eq.~\ref{eq:du}, we obtain
\begin{align}
\begin{split}
\frac{dU_2}{dz} \sim CU_1^2U_2\tau \left| \int e^{-\omega_2^2\tau^2 - i (\delta v_g^{-1}\omega_2+\delta\beta\omega_2^2)z} d\omega_2 \right|^2
\end{split}
\end{align}
This indicates that the SH pulse energy $U_2$ grows exponentially and the gain coefficient can be evaluated using normalized variables $\omega=\omega_2\tau$, normalized distance $z'=z/L$, normalized pulse bandwidth $\delta \omega = \tau^{-1} \sqrt{\delta \beta L}$, and normalized group-velocity mismatch $\Delta_v = \delta v_g^{-1} \sqrt{L/\beta}$. Then,
\begin{align}
\begin{split}
\label{eq:gamma}
\gamma(z') &= \frac{1}{U_2}\frac{dU_2}{dz'} \\
	& \sim CU_1^2 \delta \omega \left| \int e^{-\omega^2 (1-i\delta \omega^2 z') - i\Delta_v\omega\delta\omega z'}d\omega \right|^2,
\end{split}
\end{align}
where the coefficient $C$ now also includes $\sqrt{L/\beta}$. Integrating Eq.~\ref{eq:gamma}, we obtain
\begin{equation}
\label{eq:gamma2}
\gamma(z') \sim CU_1^2 \delta\omega \frac{e^\frac{-(\Delta_v\delta\omega z')^2}{2(1+\delta\omega^4z'^2)}}{\sqrt{1+\delta \omega^4z'^2}}.
\end{equation}

Taking the integral of Eq.~\ref{eq:gamma2} from 0 to 1 brings us to the final expression for the total gain as a function of both bandwidth and group-velocity mismatch, under the assumption that the fundamental pulse remains undepleted

\begin{equation}
\label{eq:gain}
G(\delta \omega, \Delta_v) \sim G_0 \delta\omega \int_0^1 \frac{e^\frac{-(\Delta_v\delta\omega z')^2}{2(1+\delta\omega^4z'^2)}}{\sqrt{1+\delta \omega^4z'^2}} dz'
\end{equation}

\begin{figure}
	\includegraphics[width=\linewidth]{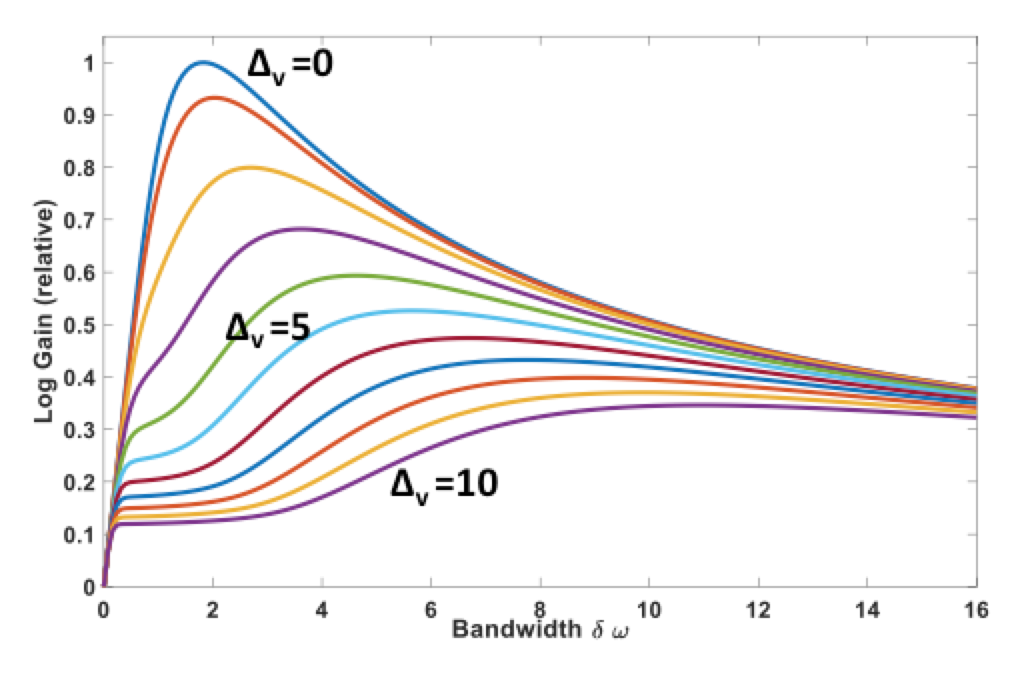}
	\caption{\label{fig:theory_gain} \textbf{The gain for second harmonic generation.} The gain for second harmonic generation as a function of the input pulse bandwidth $\delta \omega$ for various values of the normalized group velocity mismatch $\Delta_v$ demonstrates that higher gain is provided by smaller values of group-velocity mismatch. In general, when the group-velocity mismatch is higher, then smaller pulse bandwidth is required to reach higher gain. However, as the pulse bandwidth is increased to very large values, the gain increases once again, because now light is present at the frequency where group-velocity matching is achieved. Thus, for successful photo-induced SHG with femtosecond pulses, the highest conversion efficiency can be realized through perfect group-velocity matching. }
\end{figure}

For the case of no group-velocity mismatch, the solution is analytical, and rather simple
\begin{equation}
G(\delta \omega, 0) = G_0 \arcsin (\delta \omega^2)) / \delta \omega,
\end{equation}
while for finite group-velocity mismatch, we must revert to numerical integration and present the results in Fig.~\ref{fig:theory_gain}. The SHG pulse energy rises from the noise level at input $U_2 (0)$ to the output value $U_{2,\mathrm{out}}=U_2 (0)\exp(G)$. Thus, the vertical axis of Fig.~\ref{fig:theory_gain} is logarithmic and represents the output SHG energy in dB, scaled by $G_0$. If plotted on a linear scale, the curves are much sharper than they appear here and the gain for shorter pulses is much lower. For example, if one assumes $G_0=40$ dB (as in Ref.~\citenum{anderson1991}), then increasing the bandwidth from $\delta\omega = 2$ to $\delta \omega=5$ leads to the decrease of output SHG energy by 10 dB and, in the presence of waveguide loss, it is possible that no SHG signal may be detected for wider pulses.

Note that in the absence of group-velocity mismatch, the optimal gain is provided for a bandwidth $\delta\omega \approx 2$, but as the group-velocity mismatch is increased, surprisingly, the optimal bandwidth increases. This behavior results from the fact that, at very large bandwidths, light is available at the frequency where group-velocity matching is achieved. In most practical situations, higher group-velocity mismatch provides a situation where smaller bandwidth pulses can be used.

\section{Supplemental figures}

\begin{figure}
	\includegraphics[width=\linewidth]{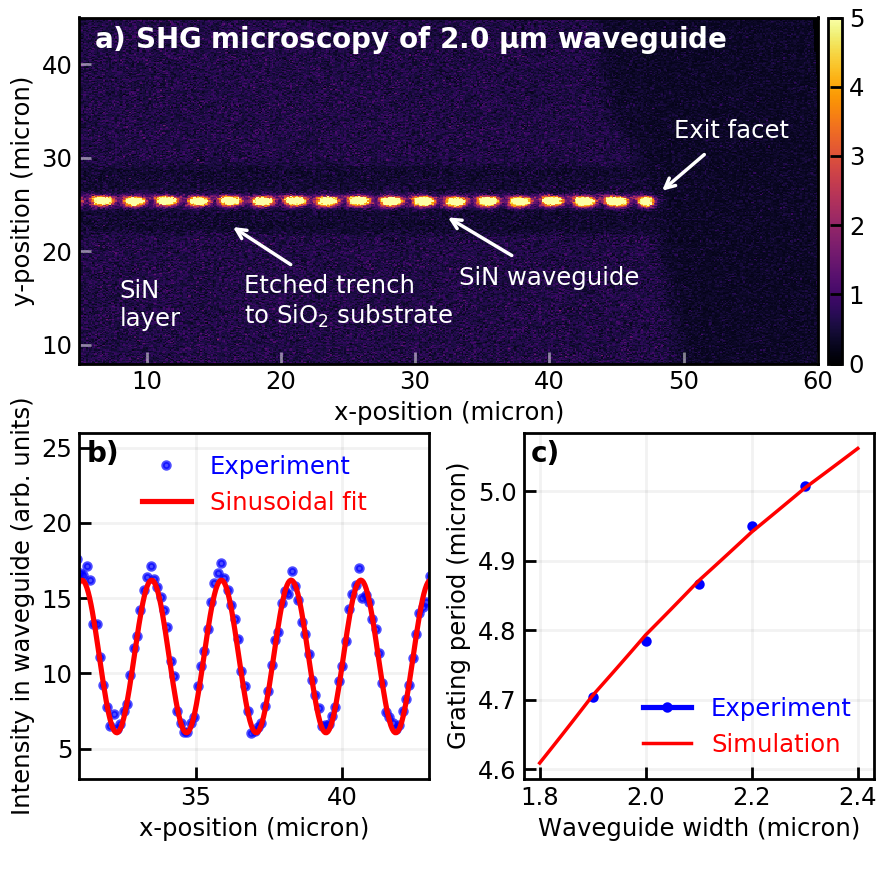}
	\caption{\label{fig:microscopy} \textbf{Microscopy of self-organized gratings.} a) An SHG-microscopy experiment reveals which regions of the sample have the strongest effective-$\chi^{(2)}$. Clear modulations can be seen, and these are only seen in the SiN waveguides that have small group-velocity mismatch. b) A simple sinusoidal function provides good agreement with observed grating. c) By fitting a sinusoidal function to all of the gratings, we can extract the grating period as a function of waveguide width. This trend closely follows the expected behavior for phase-matching SHG near 1560~nm.}
\end{figure}

\begin{figure*}
	\includegraphics[width=\linewidth]{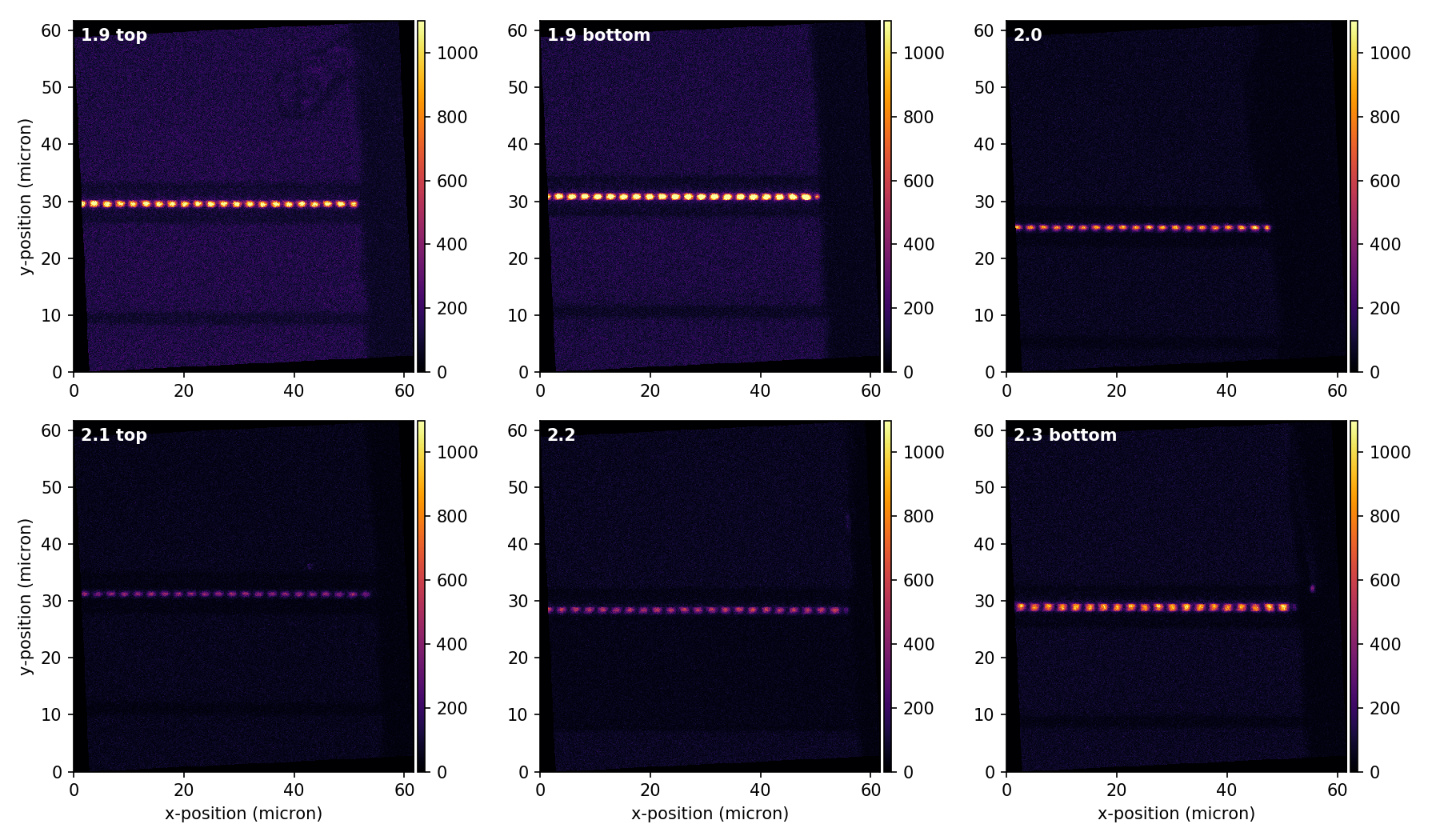}
	\caption{\label{fig:microscopy_images}
    \textbf{Second-harmonic-generation (SHG) microscopy images of the end-sections of waveguides that exhibit photo-induced SHG.} Each image corresponds to a different waveguide on the same chip. A sinusoidal modulation of the SHG intensity is produced via the interference of the fundamental and second harmonic and the period is determined by the phase-velocity mismatch between the fundamental and the second harmonic. The SHG microscopy technique is only sensitive to the magnitude of the effective $\chi^{(2)}$, and not sensitive to the sign. Consequently, the modulation period observed via the SHG microscopy experiment is $\sim$\SI{2.5}{\micro\meter}, which is half the $\sim$\SI{5}{\micro\meter} period required for quasi-phase-matching (QPM) of the SHG process. Additionally, we observe that the phase of the electric-field-induced grating at the edge of the chip is not consistent. In some cases (a and d), the grating ends at a minimum, while in other cases (b and c), the grating ends near a maximum. Thus, the phase of the grating does not appear to be fixed with respect to the end of the chip. }
\end{figure*}

\begin{figure*}
	\includegraphics[width=\linewidth]{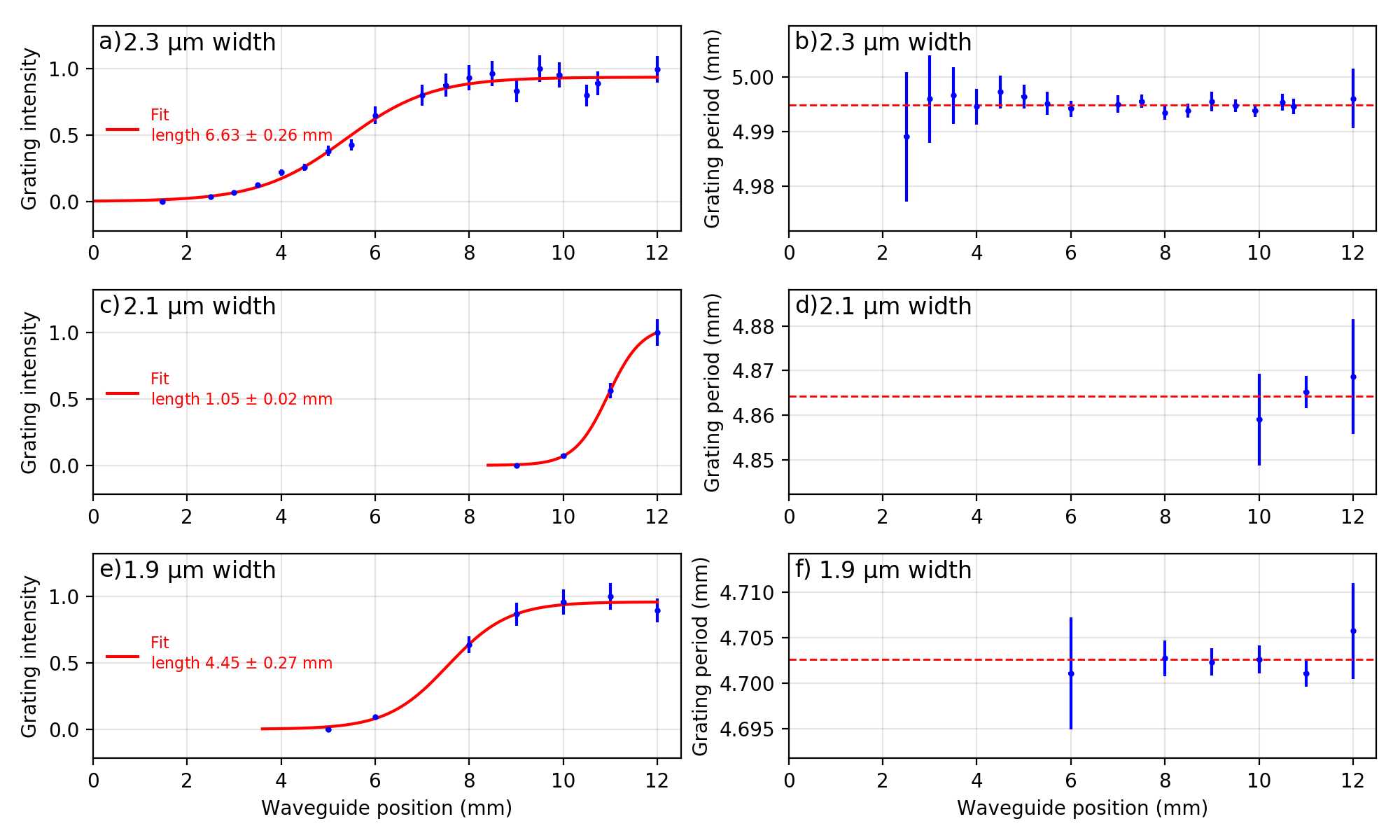}
	\caption{\label{fig:microscopy_length}
    \textbf{Estimation of the grating length.} In order to characterize the photo-induced grating on the mm length scale, numerous SHG-microscopy images were acquired along the length of the waveguide. The amplitude (left) and period (right) of the grating were extracted by fitting a sinusoidal function to vertically integrated SHG-microscopy image. The x-axis runs from the waveguide entrance to the exit, which is the direction of pulse propagation. (left) For all waveguide widths, the grating had the strongest amplitude near the exit of the waveguide and followed a sigmoidal behavior. Interestingly, the grating has significantly different lengths for different waveguide widths. (right) The grating period does not show any significant change as a function of length, within the uncertainty of this measurement. This suggests that the grating is not ``chirped'' and that a single grating period can provide QPM across the entire bandwidth of the SHG, a feature enabled via the excellent group-velocity matching provided by the high-confinement waveguide geometry.}
\end{figure*}

\begin{figure}
	\includegraphics[width=\linewidth]{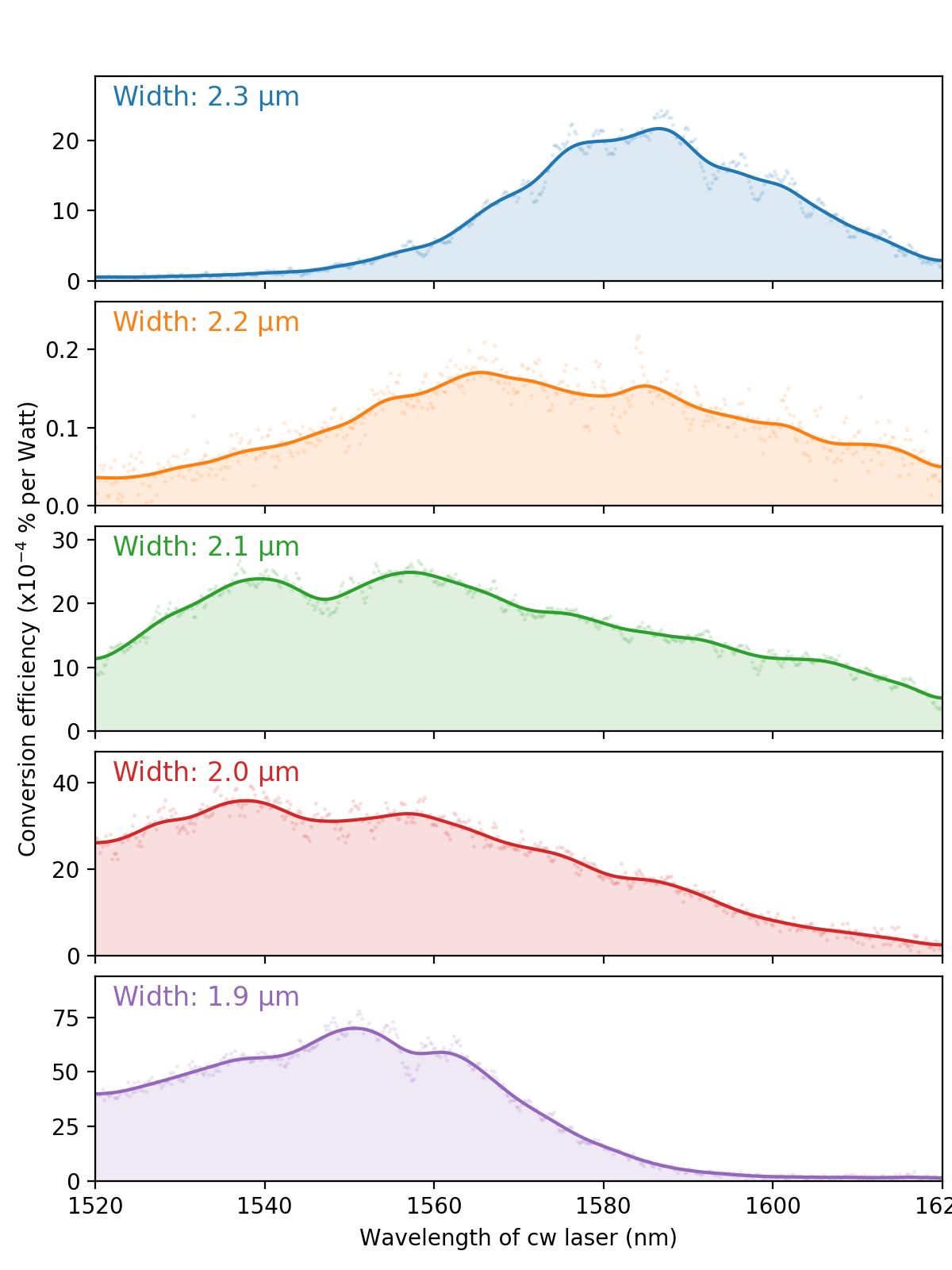}
	\caption{\label{fig:cw}
    \textbf{The photo-induced grating written by a 1560-nm femtosecond laser can subsequently be utilized for SHG of a continuous-wave (CW) laser.} Here, the CW laser was scanned from 1520 to 1620~nm and the second-harmonic power measured. The bandwidth is largest for the 2.0, 2.1, and \SI{2.2}{\micro\meter} waveguides, since these waveguides offer group-velocity matching near 1560 nm. Additionally, the wavelength that experiences the highest conversion efficiency moves to longer wavelengths with increasing waveguide width. This is likely a result of the fact that group-velocity matching causes the self-organized grating to ``select'' a certain wavelength range. Note: the line is a moving average through the data-points (dots) using a Gaussian kernel with a sigma of 10 points.}
\end{figure}

\begin{figure}
	\includegraphics[width=\linewidth]{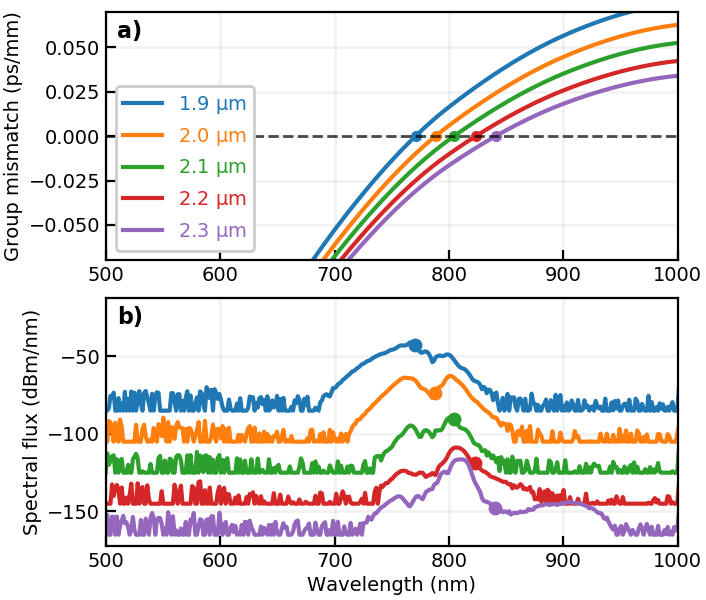}
	\caption{\label{fig:groupvelocity}
    \textbf{Group-velocity matching for SHG in silicon-nitride waveguides.} a) The wavelength of group-velocity matching for SHG can be controlled via the width of the waveguide. b) The spectrum of the second harmonic generated by femtosecond pulses moves toward longer wavelength as the waveguide width is increased. Note that the two-peaked nature of the second harmonic is a result of a two-peaked spectral shape of the pump laser.}
\end{figure}

\begin{figure*}
	\includegraphics[width=\linewidth]{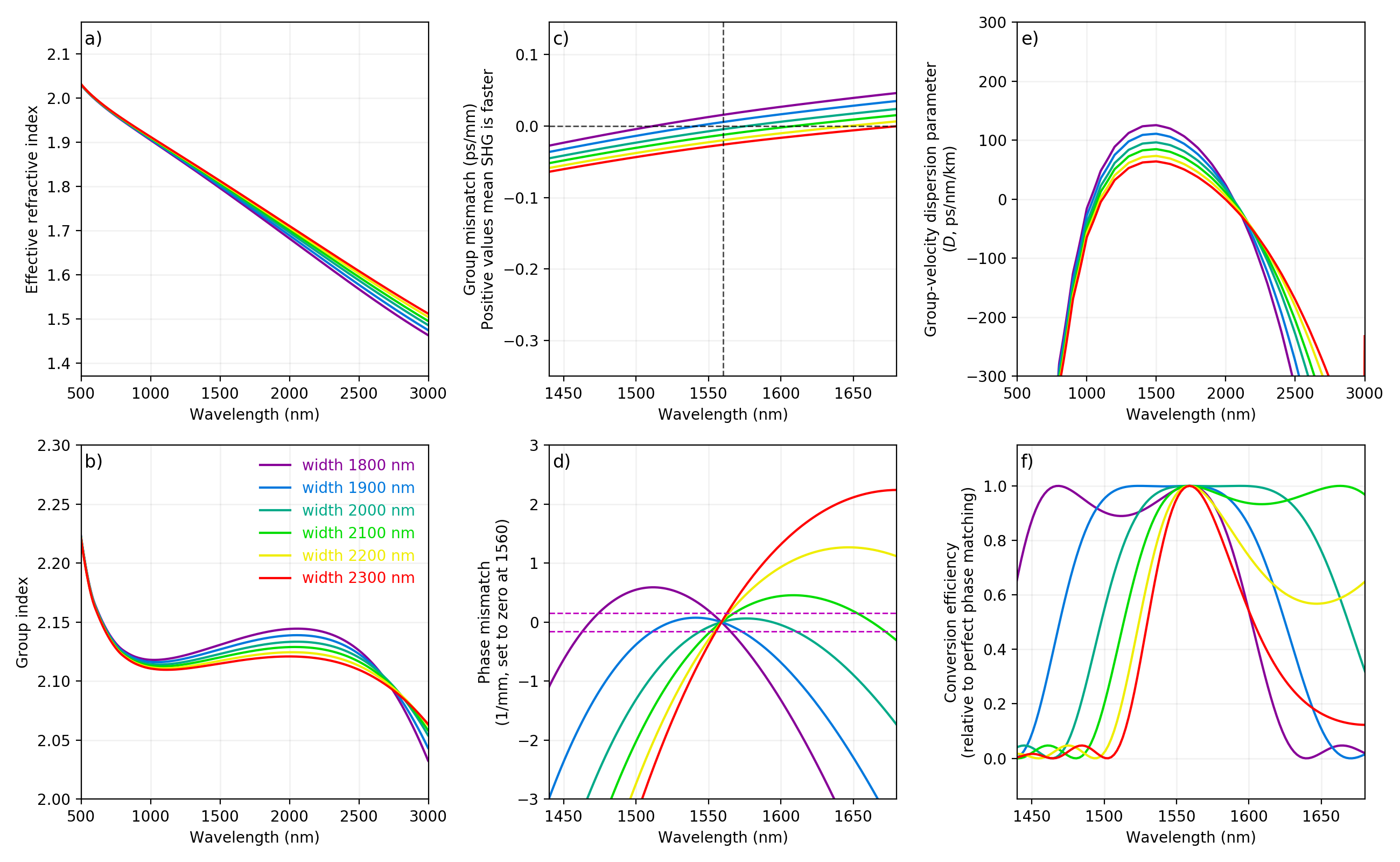}
	\caption{\label{fig:gv}
    \textbf{Calculated phase and group-velocity matching in silicon nitride waveguides.} a) The nanophotonic waveguide geometry has a small effect on the effective refractive index. b) However, the group index changes significantly, and this allows for group-velocity matching for some wavelengths. c) The group-velocity mismatch for SHG goes through zero for some waveguide widths. d) When the group-velocity mismatch is zero, then the slope of the phase-mismatch for SHG is zero at the pump wavelength, providing for broadband phase-matching. However, the phase matching isn't zero at all wavelengths, because higher order dispersion (mainly GVD) causes the phase-mismatch to increase at wavelengths far from the QPM wavelength. e) The dispersion parameter (D) is anomalous (positive) at 1560 nm. It is normal (negative) at the second harmonic. A flatter GVD curve would allow for broader-bandwidth phase matching. f) We can calculate the SHG flux as a function of wavelength given the phase-mismatch shown in the lower-center panel, and assuming a waveguide length of 6 mm. }
\end{figure*}

\begin{figure*}
	\includegraphics[width=\linewidth]{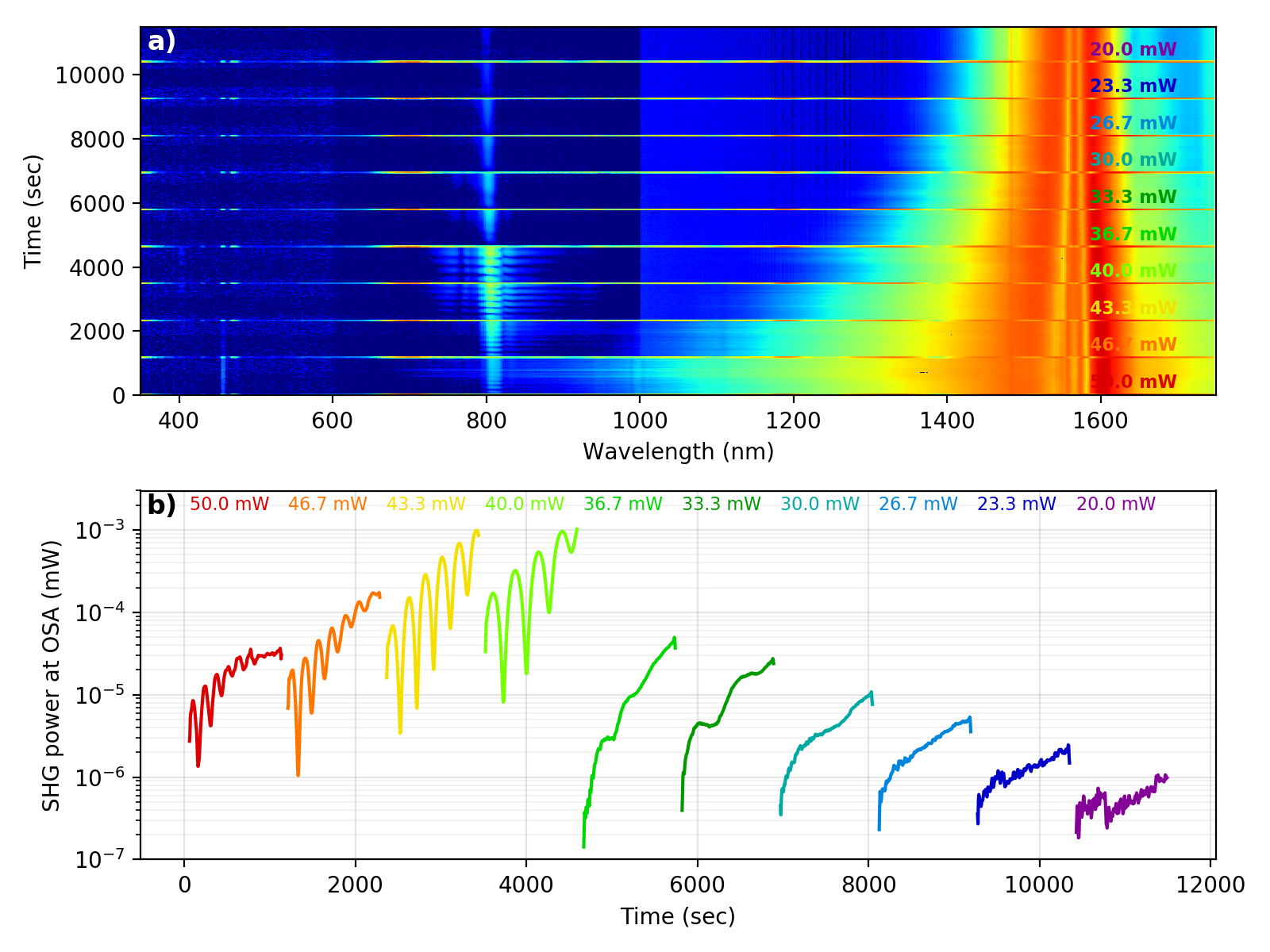}
	\caption{\label{fig:oscillations}
    \textbf{Temporal oscillations of SHG yield for the \SI{2.3}{\micro\meter}-width waveguide.} a) The spectrum of the waveguide output, recorded over several hours. During this time period, the intensity was increased to approximately 100~mW for 100 seconds, generating supercontinuum and erasing the self-organized grating. After this, the power was set to a constant level for the next 20 minutes and the second harmonic built-up. This process was repeated for numerous input power levels to investigate how the average power affects the grating formation process. Strong temporal oscillations are seen in the spectrum near 800~nm for input power levels around 40~mW. b) The integrated SHG power at different input power levels shows that, when the power level is at or above 40~mW, clear oscillations are seen in the SHG efficiency. These oscillations tend to become faster as the power level is increased, indicating some time-dependent dynamics during the self-organized grating formation process. }
\end{figure*}

\end{document}